# Roman CCS White Paper

**Title:** Optimizing Roman's High Latitude Wide Area Survey for Low Surface Brightness Astronomy

**Roman Core Community Survey:** *High Latitude Wide Area Survey*
**Scientific Categories:** *stellar populations and the interstellar medium; galaxies; large scale structure of the universe*
**Additional scientific keywords:** *Dwarf galaxies, Galaxy dark matter halos, Galaxy stellar halos, Galaxy mergers, Interacting galaxies, Galaxy clusters, Galaxy groups, Dark matter distribution.*

**Submitting Author:**
Name: Mireia Montes
Affiliation: Instituto de Astrofísica de Canarias (IAC)
Email: mireia.montes.quiles@gmail.com

**List of contributing authors** (including affiliation and email):
Francesca Annibali (INAF- Bologna, francesca.annibali@inaf.it)
Michele Bellazzini (INAF- Bologna, michele.bellazzini@inaf.it)
Alejandro S. Borlaff (NASA Ames, a.s.borlaff@nasa.gov)
Sarah Brough (U. of New South Wales, s.brough@unsw.edu.au)
Fernando Buitrago (U. of Valladolid, fbuitrago@uva.es)
Nushkia Chamba (The Oskar Klein Centre/Stockholm University, nushkia.chamba@astro.su.se)
Chris Collins (Liverpool John Moores U., C.A.Collins@ljmu.ac.uk)
Ian Dell'Antonio (Brown University, ian_dell'antonio@brown.edu)
Ivanna Escala (Carnegie Observatories/ Princeton University, iescala@carnegiescience.edu)
Anthony Gonzalez (U. of Florida, anthonyhg@astro.ufl.edu )
Benne Holwerda (U. of Louisville, bwholw01@louisville.edu)
Sugata Kaviraj (U. of Hertfordshire, s.kaviraj@herts.ac.uk)
Johan Knapen (IAC, jhk@iac.es)
Anton Koekemoer (STScI, koekemoer@stsci.edu)
Seppo Laine (Caltech/IPAC, seppo@ipac.caltech.edu)
Pamela Marcum (NASA Ames, pamela.m.marcum@nasa.gov)
Garreth Martin (Korea Astronomy and Space Science Institute/ University of Arizona garrethmartin@arizona.edu)
David Martínez-Delgado (Instituto de Astrofísica de Andalucía, dmartinez@iaa.es)
Chris Mihos (Case Western Reserve University, mihos@case.edu)
Massimo Ricotti (U. of Maryland, ricotti@astro.umd.edu)
Ignacio Trujillo (IAC, itc@iac.es)
Aaron E. Watkins (U. of Hertfordshire, a.watkins@herts.ac.uk)

*Abstract:*

One of the last remaining frontiers in optical/near-infrared observational astronomy is the low surface brightness regime (LSB, V-band surface brightness, $\mu_V \gtrsim 27$ AB mag/arcsec$^2$). These are the structures at very low stellar surface densities, largely unseen by even current wide-field surveys such as the Legacy Survey. Studying this domain promises to be transformative for our understanding of star formation in low-mass galaxies, the hierarchical assembly of galaxies and galaxy clusters, and the nature of dark matter. It is thus essential to reach depths beyond $\mu_V = 30$ AB mag/arcsec$^2$ to detect the faintest extragalactic sources, such as dwarf galaxies and the stellar halos around galaxies and within galaxy clusters. The High Latitude Wide Area Survey offers a unique opportunity to statistically study the LSB universe at unprecedented depths in the IR over an area of ≈2000 square degrees. The high spatial resolution will minimize source confusion, allowing an unbiased characterization of LSB structures, including the identification of stars in nearby LSB galaxies and globular clusters (see the white paper by Dage et al.). In addition, the combination of Roman with other upcoming deep imaging observatories (such as Rubin) will provide multi-wavelength coverage to derive photometric redshifts and infer the stellar populations of LSB objects.

## *Scientific motivation:*

Previous large-scale imaging surveys have been limited in their ability to explore the LSB universe ($\mu_V \gtrsim 27$ AB mag/arcsec$^2$). This limitation has resulted in a biased understanding of the universe, as our knowledge of the physical processes driving galaxy evolution may be significantly incomplete. The study of the LSB universe is a growing field that encompasses a wide range of topics. In this white paper, we will focus on the extragalactic LSB emission, including LSB galaxies, stellar halos, and intragroup/intracluster light.

The extragalactic LSB universe in its various forms discussed below holds the potential to provide ground-breaking insights into star formation in low-mass galaxies, the hierarchical assembly of galaxies and galaxy clusters, and the nature of dark matter. It plays a crucial role in completing our overall understanding of galaxy evolution and the physics of our universe as our theoretical models are calibrated to reproduce only a minority of objects and structures. For example, the structures found in the outer regions of galaxies not only hold valuable information about their assembly history but also have the potential to distinguish between different dark matter candidates (e.g., fuzzy dark matter). In addition, LSB galaxies represent an unexplored frontier in galaxy formation.

Progress in this field has been slow due to the scarcity of high-quality, wide-area, ultra-deep images needed to produce statistically significant and homogeneous samples. The High Latitude Wide Area Survey (HLWAS) to be conducted by the Nancy Grace Roman Space Telescope offers a unique opportunity to study these extremely faint objects[1], providing the large samples required for comprehending the mechanisms underlying LSB emission. Therefore, LSB studies and their implications for understanding our universe should be at the forefront of scientific endeavours driving the Roman mission.

### **Low Surface Brightness galaxies:**

LSB galaxies have been known to exist in abundance since the 1980s, yet we are still discovering many more of these systems. The bulk of the galaxy population, in terms of numbers (Dalcanton et al. 1997, Martin et al. 2019), is largely inaccessible to current observations. The lack of deep, wide-field surveys makes it difficult to adequately characterize the low-mass end of the galaxy luminosity function. Yet, these smaller mass scales are crucial for testing the possible solutions to several challenges to the cold dark matter cosmological paradigm (Bullock & Boylan-Kolchin 2017) and distinguish between baryonic processes and/or more exotic forms of dark matter. Furthermore, an increasing number of LSB tidal features are being discovered around nearby dwarf galaxies, providing direct evidence for their hierarchical merging assembly (e.g., Annibali et al. 2019).

Additionally, ultra-diffuse galaxies, a subset of LSB galaxies (van Dokkum et al. 2015), challenge current galaxy evolution models with their remarkable variations in dark matter and globular cluster content (e.g., van Dokkum et al. 2016, 2018, Mancera Piña et al. 2022., Gannon et al. 2022). The HLWAS will provide us with a unique opportunity to obtain the large samples needed to study these systems in detail and to discover what lies beyond the limits of our current observational capabilities.

---

[1] Surface brightness limits in the IR of $\mu_J \gtrsim 25$ AB mag/arcsec$^2$

## Baryon content in halos: from galaxies to clusters of galaxies

**Stellar Halos and Stellar Streams:**

Stellar halos are now understood to be a product of the hierarchical growth of structures (e.g., Bullock & Johnston 2005, Font et al. 2008). Galaxies undergoing accretion events leave distinct signatures in their surrounding environment, such as discrete tidal stellar streams and diffuse intrahalo light (e.g., Martínez-Delgado et al. 2010, 2023, Bell et al. 2017, D'Souza et al. 2018). Stellar halos and their substructures provide valuable insights into halo formation processes, including the connection between halo properties, merger history, and stream progenitor characteristics as well as constraints on the gravitational potential of host halos through their tidal substructures.

However, our current understanding is biased towards the information obtained from the brightest streams (the most massive progenitors) and the brightest regions of the intrahalo light (Martin et al. 2022). The depth of the HLWAS enables the exploration of the fainter intrahalo light and the detection of individual faint stellar streams in integrated light with sufficient spatial resolution to derive their properties (see Fig. 1). This allows for a direct comparison of external galaxies with the Milky Way, where faint stellar streams resulting from $10^6$ Msun galaxies are routinely identified (Helmi 2020).

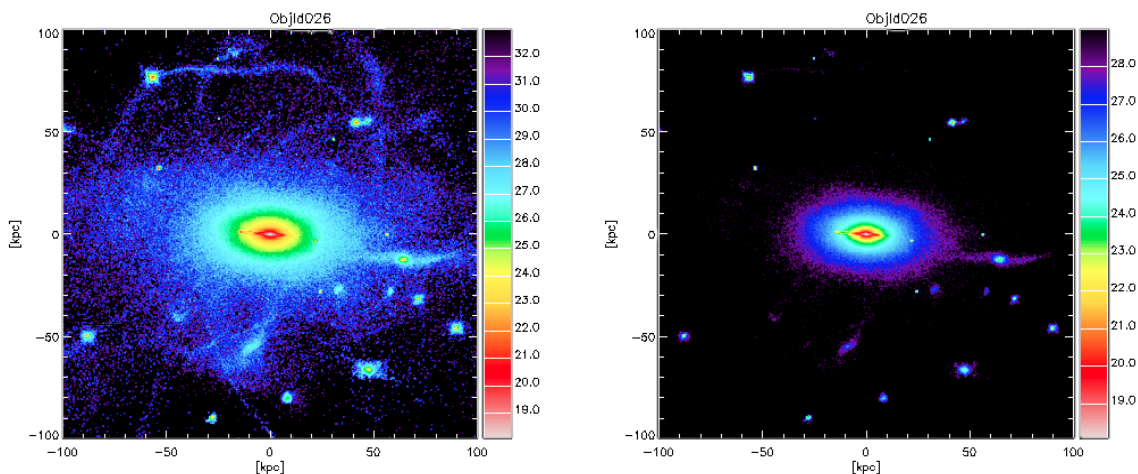

*Figure 1. A simulated galaxy seen at different depths: 33 AB mag/arcsec$^2$ (left), 29 AB mag/arcsec$^2$ (right). Reaching beyond 30 AB mag/arcsec$^2$ is crucial to witness the wealth of structures around galaxies.(Mancillas et al. 2019)*

In addition, it will allow us to explore stellar halos through star counts out to the distance of the Virgo cluster (16 Mpc), providing a more detailed view of these stellar halos (Williams et al. 2007, Smercina et al. 2023). In particular, the characterization of stellar halos around dwarf galaxies in the Local Volume has been proposed as a new method to constrain star formation in the ultra-faint dwarf galaxies in the Local Group (Kang & Ricotti 2019, Ricotti, Polisensky & Cleland 2022).

**Intragroup and Intracluster Light:**

The intracluster light (ICL, Fig. 2) is a major component of galaxy clusters preserving the history of galaxies destroyed during cluster assembly while also tracing the underlying dark matter distribution in clusters (see Mihos 2016, Montes 2022 for reviews).

Despite its importance, the lack of deep, wide-field observations providing statistically significant samples has hindered our understanding of this component, including its stellar populations and how it evolves with time. For example, in a ≈2000 square degree survey such as the HLWAS, around 100 massive clusters (>$10^{14}$ Msun) below redshift ~0.1 are expected (Mak et al. 2011). This footprint will also include a significant number of smaller groups of galaxies, as well as higher redshift clusters, where the superb spatial resolution of Roman will allow detailed studies of their diffuse light.

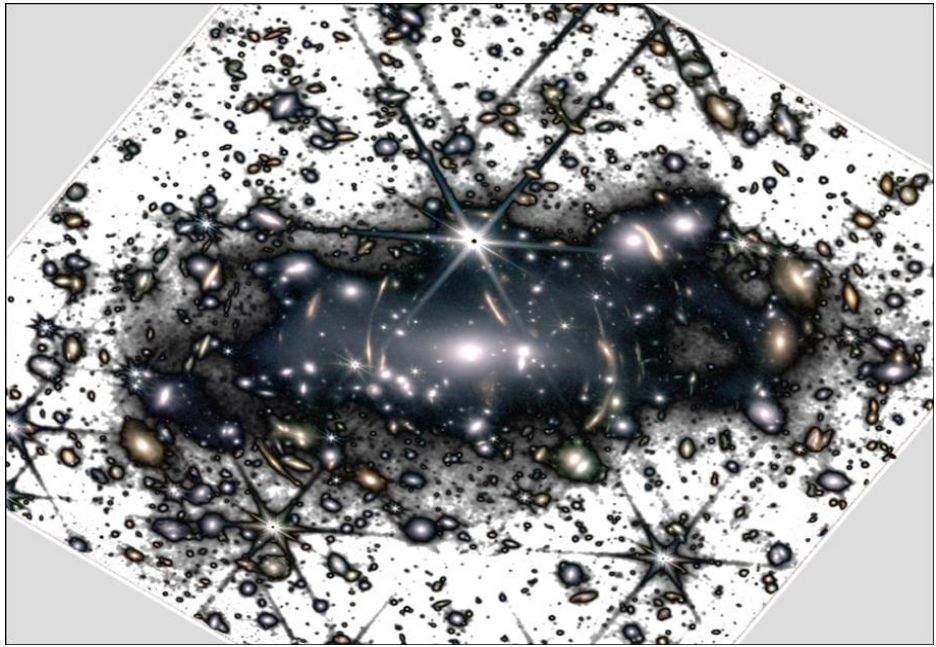

*Figure 2. JWST's IR view of the ICL of the cluster SMACS J0723.3−7327. The IR is more sensitive to the stellar populations of the ICL, making Roman the perfect instrument to study ICL at all redshifts and masses. From Montes & Trujillo (2022).*

## *Survey strategy:*

An LSB-friendly survey should aim to minimize the biases introduced by external sources (stars, cirrus) and internal sources (telescope reflections). New advances in data reduction techniques (Borlaff et al. 2019, Montes & Trujillo 2022) will take full advantage of these deep images. We summarize the key elements required to optimize Roman's HLWAS for LSB astronomy below.

**Image depth:**
The expected 5σ point source AB magnitude of the HLWAS is ~26.5 mag. This corresponds to surface brightness limits (3σ in a 10x10 arcsec² area) of 30.5 mag/arcsec², which represents unprecedented depths for an IR survey covering ≈2000 square degrees²[2]. By comparing it with existing IR LSB studies, we will be able to study the ICL at distances greater than 200 kpc up to a redshift of ~0.4 (Montes & Trujillo, 2022), as well as stellar halos at distances greater than 20 kpc up to a redshift of 1 (Borlaff et al., 2019). These observations will allow us to recover complete samples of LSB galaxies with masses well below $10^9$ Msun at distances beyond the Local Group (Martin et al., 2019).

---

[2] For comparison, the IR Hubble Ultra Deep field reaches depths of 32.5 mag/arcsec² (Borlaff et al. 2019) but only covers 11 arcmin², limiting its impact for statistical studies.

It is important to note that very short exposure times will increase the noise (readout noise) in the images, resulting in a final coadd that is shallower than expected. Therefore, we request single pointing exposure times on the order of hundreds of seconds to mitigate this issue.

**Survey location:**

Sources of contamination are zodiacal light, Galactic cirrus (filaments of dust of our Galaxy, Román et al. 2020) and stars. Consequently, we want to avoid the ecliptic and the Galactic plane, including areas that are significantly affected by cirrus contamination. By doing so, we can reduce the impact of these contaminants on the survey data.

Maximizing the overlap between the HLWAS and other surveys such as Rubin (Gezari et al., 2022) will enhance the scientific value of the HLWAS. This collaboration will provide increased wavelength coverage and, in turn, benefit Rubin by leveraging the exquisite spatial resolution offered by observations from space. Having both deep optical and deep IR images will be critical for obtaining reliable physical parameters such as photometric redshifts and stellar masses for faint galaxies. By combining the strengths of these surveys, we can achieve a more comprehensive and valuable understanding of the sources.

**Observing Strategy:**

Reflections in the optical path are a real concern for deep imaging. They can be mitigated by dithering, so that they do not fall in the same region of the detector. Typical dithering strategies that account for missed regions (detector gaps, cosmic rays, detector defects…) are too small to minimize reflections. For this, we would ideally need a shift as large as the source we are studying (e.g., Trujillo & Fliri 2016). We propose a minimum dithering shift of 1 arcmin for the HLWAS survey which is approximately the size of a Milky Way-like galaxy at 50 Mpc. By implementing this dithering strategy, we strike a balance between maintaining the survey's depth and ensuring the dithering is sufficiently useful for most science cases outlined in this white paper. Together with the two roll angles expected (Troxel et al. 2021), this dithering strategy will help minimize scattered light from the telescope maximizing the depth the survey will reach.

**Filter choice:**

For LSB studies, we are interested in broad filters with sufficient coverage to sample the SED for photometric redshifts and stellar population analysis. The currently proposed set Y106/J129/H158/F184 (WFIRST-SYS-REQ-0020, Troxel et al. 2021), spanning the range from 0.93-2.00 μm, will be ideal for this.